\documentclass[12pt]{article}

\usepackage{graphicx}

\newcommand{\beq}{\begin{equation}}
\newcommand{\eeq}{\end{equation}}
\newcommand{\beqa}{\begin{eqnarray}}
\newcommand{\eeqa}{\end{eqnarray}}
\newcommand{\beqar}{\begin{eqnarray*}}
\newcommand{\eeqar}{\end{eqnarray*}}

\begin{document}
\thispagestyle{empty}


\vspace{32pt}
\begin{center}

\textbf{\Large Cosmic Magnetic Lenses}

\vspace{40pt}

Eduardo Battaner, Joaqu\'\i n Castellano, Manuel Masip
\vspace{12pt}

\textit{Departamento de F{\'\i}sica Te\'orica y del Cosmos}\\
\textit{Universidad de Granada, E-18071 Granada, Spain}\\
\vspace{16pt}
\texttt{battaner@ugr.es, jcastellano@ugr.es, masip@ugr.es}

\end{center}

\vspace{30pt}

\date{\today}

\begin{abstract}

Magnetic fields play a critical 
role in the propagation of charged cosmic rays. Particular 
field configurations supported by different astrophysical
objects may be observable in cosmic ray maps.
We consider a simple configuration,
a constant azimuthal field in a disk-like object, that we
identify as a {\it cosmic magnetic lens}. Such configuration
is typical in most spiral galaxies, and we assume that it can
also appear at smaller or larger scales.
We show that the magnetic lens deflects
cosmic rays in a regular geometrical pattern,
very much like a gravitational lens deflects light but
with some interesting differences. In particular, 
the lens acts effectively only
in a definite region of the cosmic-ray spectrum, and
it can be convergent or divergent depending on the
(clockwise or counterclockwise) direction of the magnetic field
and the (positive or negative) electric charge of the cosmic ray.
We find that the image of a point-like
monochromatic source may be one, two
or four points depending on the relative position of source, observer and
center of the lens. For a perfect alignment and a lens in the orthogonal
plane the image becomes a ring.
We also show that the presence of a lens 
could introduce low-scale fluctuations and 
matter-antimatter asymmetries in the fluxes from distant sources.
The concept of cosmic magnetic lens that we introduce here may be
useful in the interpretation of possible patterns 
observed in the cosmic ray flux at different energies.

\end{abstract}

\newpage

\section{Introduction}

High-energy cosmic rays carry information from their source and
from the medium where they have propagated in their way to the
Earth. They may be charged particles (protons, nuclei or
electrons) or neutral (photons and neutrinos). The main difference
between these two types of astroparticles is that the first one
loses directionality through interactions with galactic and
intergalactic magnetic fields. In particular, random background
fields of order $B\approx 1$ $\mu$G in our galaxy will uncorrelate
a particle from its source after a distance larger than
\begin{equation}
r_g={E\over e c B}\approx {E\over 1\;{\rm TeV}} \times
10^{-3}\;{\rm pc}\;, 
\label{rg} 
\end{equation}
where $e$ is the unit charge
and $E$ the energy of the particle. As $E$ grows the reach of
charged particles increases, extending the distance where they may
be used as astrophysical probes. At $E\approx 10^9$ GeV this
distance becomes 1 Mpc, and cosmic rays may bring information from
an extragalactic source.
Of course, it seems difficult to
imagine a situation where charged cosmic rays may be used to {\it
reveal} or characterize an object. In this letter we propose that
they can detect the presence of an astophysical object, {\it
invisible} to high-energy photons and neutrinos, that we name as
{\it cosmic magnetic lens} (CML).

The term {\it magnetic lensing} has already been used
in the astrophysical literature to describe, generically,
the curved path of charged cosmic rays through a magnetized medium.
Harari et al. \cite{harari2001,harari2005,harari2010} 
studied the effect of galactic
fields, showing that they may produce
magnification, angular clustering and caustics.
Dolag et al. \cite{dolag2009} considered lensing by the tangled
field of the Virgo cluster, assuming that the galaxy M87
was the single source of ultrahigh energy cosmic rays.
Shaviv et al. \cite{shaviv1999} studied the lensing near
ultramagnetized neutron stars.
Our point of view, however, is different. 
The CML will be defined by a basic magnetic-field configuration
with axial symmetry that could appear in astrophysical objects
at any scale: from clusters
of galaxies to planetary systems. 
The effect of the CML on galactic cosmic rays 
({\it i.e.}, charged particles of energy $E<10^9$ GeV) 
will not be significantly altered
by turbulent magnetic fields if the lens is 
within the distance $r_g$
in Eq.~(\ref{rg}) and its magnetic field is substantially
stronger than the average background field between its position
and the Earth. Since the CML is 
a definite object, we can separate source, magnetic lens
and observer. Although it is not a lens in the geometrical
optics sense (the CML does not have a focus), its effects
are generic and easy to parametrize, analogous to
the ones derived from a gravitational lens (with no focus
neither).

\section{A magnetic lens}

The basic configuration that we will consider is an azimuthal mean
field $\vec B$ in a disc of radius $R$ and thickness $D$. The
field lines are then circles of radius $\rho\le R$ around the disk
axis. As a first approximation we will take a constant intensity $B$, 
neglecting any dependence on $\rho$ (notice, however, that
a more realistic $B$ should vanish smoothly at
$\rho=0$ and be continuous at $\rho=R$). 
Our assumption will simplify the
analysis while providing all the main effects of a magnetic lens.
The disk of most spiral galaxies has a large toroidal
component of this type \cite{beck2005}, 
so they are obvious candidates to CML. The
configuration describing the CML would be natural wherever there
is ionized gas in a region with turbulence,
differential rotation and axial symmetry,
since in such environment the magnetic field tends to be amplified
by the {\it dynamo effect} \cite{parker1971,brandenburg2005}. 
We will then assume that it may appear
at any scale $R$ with an arbitrary value of $B$.

\begin{figure}
\begin{center}
\includegraphics[width=7.6cm]{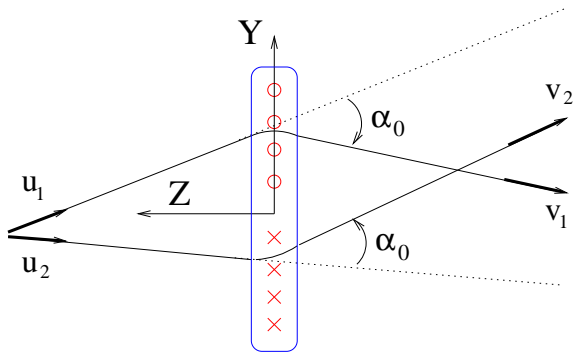}
\end{center}
\caption{
Trajectories in the $x=0$ plane. $\vec B \propto (1,0,0)$ at $y>0$ and 
$\vec B \propto (-1,0,0)$ at $y<0$.
}
\end{figure}

Let us parametrize the magnetic field and its effect on a charged
cosmic ray. If the lens lies in the $XY$ plane with the center
at the origin (see Fig.~1) $\vec B$ 
is\footnote{A continuous field configuration could be
modelled just by adding a factor of
$\left( 1-\exp \left[ \left( \rho/\rho_0 \right)^{n_0} \right]
\right) \times
\exp\left[ \left(\rho/R \right)^{n_R} \right] \times
\exp\left[ \left(2z/D \right)^{n_D} \right]$. When the integers
$n_0$, $n_R$ and $n_D$ are chosen very large and $\rho_0$ very
small we recover our disc
with a null $B$ at $\rho=0$.}

\begin{equation}
\vec B = \left\{
\begin{array}{l l}
\displaystyle
\displaystyle {B\over \rho} \left( y, -x, 0 \right)\;\;\;\;
& \displaystyle
{\rm if}\;\;\rho < R\;\;{\rm and}\;\; |z| < {D\over 2}\;;\\
\\
\displaystyle
0 & {\rm otherwise}\;,
\end{array} \right.
\label{B} 
\end{equation}
with $\rho\equiv \sqrt{x^2+y^2}$. To understand its
effect, we will first consider a particle moving
in the $YZ$ ($x=0$) plane with direction $\vec u$
(the case depicted in Fig.~1).
When it enters the lens the cosmic ray finds an orthogonal
magnetic field that curves its trajectory. The particle then
rotates clockwise\footnote{We define a positive deviation
$\alpha_0$ if the rotation from $\vec u$ to $\vec v$ around the
axis $\vec u_B$ is clockwise.} around the axis $\vec u_B=\vec
B/B$, describing a circle of gyroradius $r_g=E/(ecB)$. The segment
of the trajectory inside the lens has a length $l\approx D$, so
the total rotation angle $\alpha_0$  when it departs is 
\begin{equation}
\alpha_0\approx {e c B D\over E}\;. \label{alpha0} 
\end{equation}
The direction of the particle after
crossing the lens is then $\vec v =
R_B(\alpha_0) \;\vec u$. The angle $\alpha_0$ will be 
the only parameter required to describe the effect of this
basic lens. An important point is that $\vec B$ and
the Lorentz force change sign if the trajectory goes through
$y<0$. In that case the deflection is equal in modulus but
opposite to the one experience by particles going through $y>0$
(see Fig.~1). Therefore, the effect of this lens is {\it
convergent}, all trajectories are deflected the same angle
$\alpha_0$ towards the axis of the lens. Notice that the lens
changes to {\it divergent} for particles of opposite electric
charge or for particles reaching the lens from the
opposite ($z<0$) side.

The effect on a generic trajectory within a plane not
necessarily orthogonal to the lens is a bit more involved.
It is convenient to separate
\begin{equation}
\vec u = \vec u_\parallel + \vec u_\perp\;;\;\;\;
\vec v = \vec v_\parallel + \vec v_\perp\;,
\end{equation}
where $\vec u_\parallel=(\vec u \cdot \vec u_B)\; \vec u_B$
and $\vec u_\perp = \vec u -\vec u_\parallel $ are parallel and
orthogonal to the magnetic field, respectively (and analogously
for $\vec v$).
In this case the magnetic field will rotate the initial
direction $\vec u$ an
angle of $\alpha=u_\perp\alpha_0$ around the
axis $\vec u_B$:  $\vec v=R_B(u_\perp\alpha)\;  \vec u$.
This means that the parallel components of the initial
and the final directions coincide,
\begin{equation}
u_\parallel= \vec u \cdot \vec u_B = v_\parallel\;,
\label{parallel}
\end{equation}
whereas the orthogonal component $\vec u_\perp$, of modulus
$u_\perp=\sqrt{1-(\vec u \cdot \vec u_B)^2}$, rotates into
\begin{equation}
\vec v_\perp = \cos (u_\perp \alpha_0)\; \vec u_\perp
- \sin (u_\perp \alpha_0)\; \vec u_B\times \vec u_\perp\;.
\label{perp}
\end{equation}

An important observation concerns the {\it chromatic aberration}
of the lens. The deviation $\alpha_0$ caused
by a given CML is proportional to the inverse energy of the
cosmic ray. If $E$ is small and $\alpha_0> \pi/2$, then the
lens acts {\it randomly} on charged particles, diffusing them
in all directions. On the other hand, if $E$ is large the
deviation becomes small and is smeared out as the particle
propagates to the Earth. Only a region of the cosmic-ray
spectrum can {\it see} the CML.

\section{Image of a point-like source}

Let us now study the image of a localized monochromatic 
source produced by the CML. We will
consider a {\it thin} lens $(R\gg D)$ located on the
plane $z=0$ (see Fig.~2). Its effect on a cosmic ray can be
parametrized in terms of the angle $\alpha_0$ given in
Eq.~(\ref{alpha0}). The rotation axis is
\begin{equation}
\vec u_B={1\over \sqrt{x^2+y^2}} (y,-x,0)\;,
\end{equation}
and the coordinates of source and observer
are $S=(s_1,s_2,s_3)$ and $O=(o_1,o_2,o_3)$, respectively.
We will use the axial symmetry of the lens to set
$s_1=0$.
\begin{figure}
\begin{center}
\includegraphics[width=8.cm]{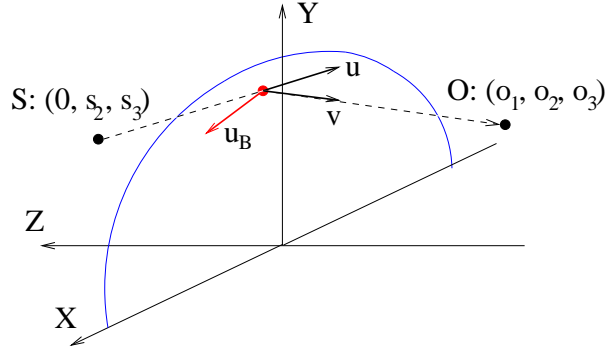}
\end{center}
\caption{Trajectory from the source to the observer.
}
\end{figure}
The trajectory will intersect the CML at
$(x,y,0)$. There the
initial direction $\vec u$ will change to $\vec v$, with
\begin{equation}
\vec u={(x,y-s_2,-s_3)\over \sqrt{x^2+(y-s_2)^2+s_3^2}}
\;,\;\;\;
\vec v={(o_1-x,o_2-y,o_3)\over \sqrt{(o_1-x)^2+(o_2-y)^2+o_3^2}}
\;.
\label{u-v}
\end{equation}

\begin{figure}
\begin{center}
\includegraphics[width=8.cm]{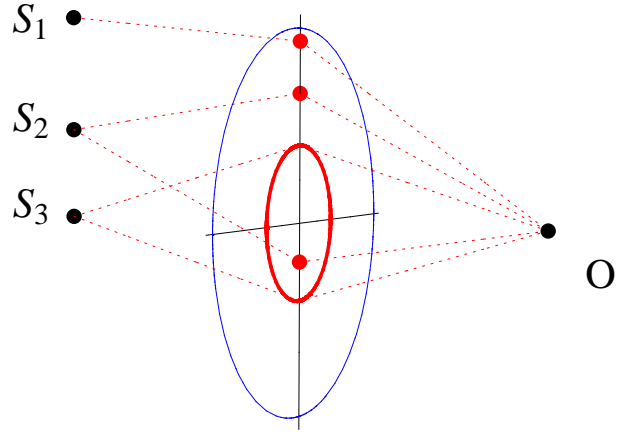} 
\end{center}
\caption{Trajectories with $\beta>\alpha$ ($S_1$),
$\beta<\alpha$ ($S_2$) and $\beta=0$ ($S_3$) for an observer at
the axis.
\label{fig3}}
\end{figure}
Therefore, given a source $S$, an observer $O$ and a lens
producing a deviation $\alpha_0$, we can determine the
coordinates $(x,y,0)$ where the rotation $R_B(u_\perp \alpha)$ 
described in the previous section exactly transforms 
$\vec u$ into $\vec v$. The first
condition on $x$ and $y$,
given in Eq.~(\ref{parallel}), is that $\vec B$ does not
change the longitudinal component of the velocity,
\begin{equation}
\vec u \cdot \vec u_B = \vec v \cdot \vec u_B \;.
\end{equation}
The second one, derived from Eq.~(\ref{perp}), defines the
rotation of $\vec u_\perp$ produced
by the magnetic field. It can be written 
($u_\perp=|\sin \widehat{\vec u \vec B} |$)
\begin{eqnarray}
{\vec v_\perp \cdot \vec u_\perp\over u_\perp^2}
& = & \cos\; (u_\perp \alpha_0) \;, \cr
{\vec v_\perp \cdot (\vec u_\perp\times \vec u_B)
\over u_\perp^2}
& = & \sin\; (u_\perp \alpha_0) \;.
\end{eqnarray}
The second equation above is necessary to fully specify
the rotation. Notice that $\alpha= u_\perp \alpha_0$
has a definite sign: positive for a convergent CML and
negative for a divergent one. In addition, the solution
must verify that $x^2+y^2<R^2$.

We find that for $R\rightarrow \infty$ and a convergent
lens there is always at
least one solution, whereas for a divergent one there
is a region around the axis that my be {\it hidden} by
the CML (this region disappears if $B$ goes smoothly to
zero at the center of the lens).
To illustrate the different possibilities in Fig.~3--left
we
have placed the observer in the axis at a distance $L$ 
from the lens, $O=(0,0,-L)$,
and have parametrized the position of the source 
(at a distance $d$ from the lens) as
$S=(0,d \sin\beta, d\cos\beta)$. In this
case $u_\parallel=0=v_\parallel$ and $u_\perp=1$. If
the lens is convergent ($\alpha_0>0$) and
$|\beta|>\alpha_0$, then the image of the source is just
a single point. For a source at $|\beta|<\alpha_0$ we
obtain two solutions, which correspond to trajectories
from above or below the center of the lens. 
For
a source in the axis ($\beta=0$) the solution is a
ring of radius
\begin{equation}
r={d+L\over 2\tan\alpha_0} \left(
\sqrt{1+{4 d L \tan^2\alpha_0\over (d+L)^2}}-1 \right)
\;.
\label{ring}
\end{equation}

If the observer is located out of the axis
but still in the $x=0$ plane the possibilities
are similar, but the ring becomes a {\it cross} similar to
the one obtained through gravitational lensing.
Finally, if we take the observer out of the $x=0$ plane there
appears always a single solution.

\section{Fluxes from distant sources}

Let us finally explore how the presence of a CML
changes the flux $F$ 
of charged particles from a localized source $S$.
It is instructive to consider 
the case where $S$ is 
a homogeneous disk of radius $R_S$ placed 
at a distance $d$ from the lens and 
the observer $O$ is at a large distance $L$, 
\begin{equation}
R_s < d,R \ll L\,,
\label{f1}
\end{equation}
as shown in Fig.~4.
In addition, we will assume that the magnetic field defining the
lens goes smoothly to zero near the axis, and that the
source is monochromatic.

If there were no lens, $O$ would see $S$ under
a solid angle 
\begin{equation}
\Delta \Omega_0 \approx \pi {R_S^2\over L^2}\,.
\end{equation}
If all the points on $S$ are equally bright and
the emission is isotropic, the differential flux 
${\rm d}F / {\rm d}\Omega$ from all the directions inside
the cone $\Delta \Omega_0$ will be approximately the same, 
implying a total
flux (number of particles per unit area)
 \begin{equation}
F_0 = \int_{\Delta \Omega_0} {\rm d} \Omega\,
{{\rm d} F \over {\rm d} \Omega}\approx 
\pi {R_S^2\over L^2}\,{{\rm d} F \over {\rm d} \Omega}\,.
\end{equation}

The lens in front of $S$ will deflect
an approximate angle 
$\alpha$ all trajectories crossing far from the
axis. In Fig.~4 we have pictured\footnote{A pointlike source
in the axis is transformed by the lens into a ring, as explained
in Section 4. As the source grows, the ring becomes thicker and 
eventually closes to a circle, which is the case considered in
Fig.~4.} the limiting directions
reaching the observer, that define a cone 
\begin{equation}
\Delta \Omega_+ \approx \pi {(R_S+d\, \tan\alpha)^2\over L^2}\,.
\end{equation}
$O$ sees now cosmic rays from directions inside the larger cone
$\Delta \Omega_+$ or, in other words, sees the radius $R_S$ of
the source {\it amplified} to $R_S+d\, \tan\alpha$.
\begin{figure}
\begin{center}
\includegraphics[width=9.5cm]{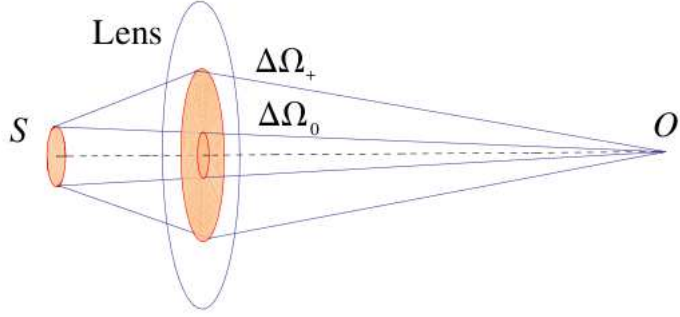} 
\end{center}
\caption{Cone of trajectories from $S$ to $O$ 
with and without lens for a homogeneous and
monochromatic source.
\label{fig4}}
\end{figure}

We can then use Liouville's theorem to deduce how the flux 
observed by $O$ is affected by the presence of the lens.
This theorem, first applied to cosmic rays moving inside 
a magnetic field by Lemaitre and Vallarta \cite{lemaitre1933}, 
implies that an observer following a trajectory 
will always observe the same differential flux (or intensity,
particles per unit area and solid angle) along the
direction defined by that trajectory. 
For example, in the case with no lens an observer in the
axis at a distance $L'\gg L$ will still observe the same
differential flux ${\rm d} F / {\rm d} \Omega$. However,
the cone of directions that he sees will be smaller, 
$\Delta \Omega'_0\approx \pi \,R_S^2/ L'^2$, and the total
flux from that source will scale like $F'\approx F\,L^2/L'^2$.
The effect of the lens is then just to change the cone of directions
reaching $O$ from $S$, without changing the differential flux. 
This implies an integrated flux
\begin{equation}
F_+\approx F_0 {\Delta \Omega_+\over \Delta \Omega_0} 
\approx  F_0 \left( 1+{d^2\, \tan^2\alpha\over R_S^2}\right)\,.
\end{equation}
An important point is that the solid angle intervals 
$\Delta \Omega_{0,+}$ will in general be much smaller than the angular
resolution at $O$. As a consequence, an observer trying to 
measure a differential flux will always include the whole cone
$\Delta \Omega_{0,+}$ within the same solid angle bin: only 
the integrated fluxes $F_{0,+}$ (averaged over the angular resolution)
are observable. 

Now let us suppose that there are many similar sources at approximately
the same distance from the observer and covering a certain range of 
directions. Cosmic rays emitted from each source 
will reach $O$ within a very
tiny cone $\Delta \Omega_{0}$, and will be observed integrated over
that cone and averaged over the angular resolution. If one of the sources
has in front a CML, its cone $\Delta \Omega_{+}$ at $O$ and thus 
its contribution to one of the direction bins 
will be larger, what would translate
into a low-scale anisotropy\footnote{The direction of the source would
be measured with a gaussian distribution that could 
take it to adjacent bins.} 
within the range of directions covered by the sources 
(see Fig.~5, left). 
\begin{figure}
\begin{center}
\includegraphics[width=8.5cm]{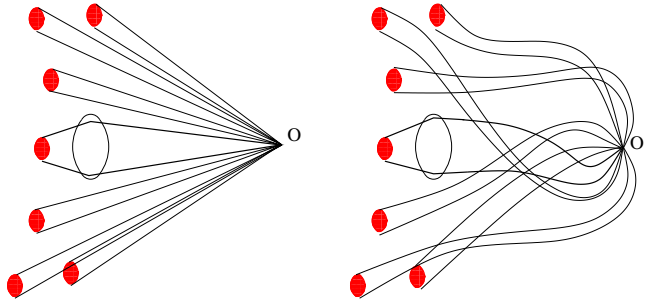} 
\end{center}
\caption{Trajectories from $S$ to $O$ without (left) 
and with (right) irregular magnetic fields along the trajectory.
\label{fig5}}
\end{figure}

In principle, this effect would not be erased by
irregular magnetic fields from the source to the observer, 
that deflect the trajectories
and tend to {\it isotropize} the fluxes (in Fig.~5, right).
The contribution 
from the source behind the CML (reaching now $O$ from a different
direction) will
still tend to be larger. The effect of the lens
is to increase the {\it size} $R_S$ of the source to 
$R_S+ d\, \tan\alpha$; random magnetic fields 
will change the direction of arrival and the 
effective distance between $S$ and $O$ ({\it i.e.}, the direction
and the size of the cone from each source), 
but not the initial deflection produced by the lens
nor (by Liouville's theorem)
the differential flux within each tiny cone. Therefore,
the cone from the source behind the lens tends to be larger, 
and when integrated and averaged over the resolution bin 
may still introduce a low-scale
anisotropy. The effect, however, tends to vanish if the cones 
are so small that the probability to observe two particles from 
the same cone of directions is smaller than the probability to 
observe particles from two disconnected cones with origin in 
the same source ({\it i.e.}, 
in the deep diffuse regime where trajectories become random walks).

Finally, note that the effect of a divergent CML would be just
the opposite. The presence of a lens 
could then introduce an excess for
positive charged particles and a deffect for the negative ones
(or a matter--antimatter asymmetry if both species were
equally emitted by $S$). 

\section{Summary and discussion}
It is known that galactic and intergalactic
magnetic fields play a very important role
in the propagation of charged cosmic rays. Here we have
explored the effect of a very simple configuration,
a constant azimuthal field in a thin disk,
that we identify as a CML. Such object acts on cosmic
rays {\it like}
a gravitational lens on photons, with some very
interesting differences. Gravitational lenses are 
always convergent, whereas if a 
magnetic lens is 
convergent for protons and positrons, it changes to
divergent for antiprotons and electrons. In addition,
the deflection that the CML produces depends on the particle
energy, so the lense is only visible in a very definite
region (around one decade of energy) of the spectrum.

Our intention has been to introduce the concept of CML
and discuss its possible effects leaving the
search for possible candidates for future work.
Generically,
the magnetic-field configuration
defining the CML is {\it natural} and tends to be 
established by the
dynamo effect.
For example, in spiral galaxies $\vec B$
can be pure azimuthal (the one we have assumed), 
axisymmetric spiral or bisymmetric spiral, with or without
reversals \cite{beck2005,battaner2008}, 
but in all
cases the azimuthal component dominates. 
Our galaxy is  not an  exception
\cite{han2009,ruiz2009}, it includes
in the disk a spiral magnetic field
of $B\approx 4$ $\mu$G. This would actually force that
any analysis of magnetic lensing by other
galaxies must {\it subtract}
the effect produced by our own magnetic field.
CMLs could also be present in galactic halos, as
there are observations of polarized synchrotron emission
suggesting the presence of regular fields \cite{dettmar2006}.
Analogous indications \cite{bonafede2009}
can be found for
larger structures, like clusters and their halos.
Inside our galaxy, the antisymmetric tori placed 1.5 kpc away
in both hemispheres discovered by Han et al. 
\cite{han1997} would also produce magnetic lensing
on ultrahigh energy cosmic rays.
At lower scales (20--800 pc) molecular clouds and 
HII regions \cite{gonzalez1997}
are also potential candidates.
Molecular clouds have strong regular fields in the 
range of $0.1$--$3$ mG \cite{crutcher2005}.
Moreover, many reversals in the field direction observed 
in our galaxy seem to coincide with HII regions
\cite{wielebinski2005}, which 
would indicate that the field follows the rotation
velocity in that region.  
There are also observations of Faraday screens covering  
angles of 
a few minutes of unknown origin \cite{mitra2003}.
Finally, nearby protostellar disks may provide a
magnetic 
analogous of the gravitational microlenses, as they
define small objects of $\approx 10^3$ AU diameter
with azimuthal magnetic fields 
\cite{stepinski1995} of order 
tens of mG \cite{goncalves2008}.
Therefore, we think it is justified to presume that
CMLs may appear at
any scales $R$ with different values of $B$.

The lensing produced by a CML will be 
affected by the turbulent magnetic fields, but under certain 
conditions they should
remain observable. For example, the typical 
lensing produced by a galaxy
on cosmic rays of energy above $10^9$ GeV is caused by a
regular magnetic field of order $\mu$G, while the distortions
will come from  fluctuations of the same order.
The region of coherence of these magnetic fluctuations, however, 
is just around 10--100 pc, varying randomly
from cell to cell. Since the regular field that define the lens
will act along distances 10--100 times larger, its effect 
on cosmic rays will dominate,
and the expected {\it blurring} due to turbulences will be small.
For CMLs inside our galaxy one should in general {\it subtract} 
the effect due to the local field at the relevant scale.
Suppose, for example, that we have
a small lens ($D\approx 10^{-3}$ pc) with a strong 
magnetic field ($B\approx 1$ mG) at a distance below 10 pc
from the Earth. If the magnetic field along the
 trajectory from the lens to the Earth 
is of order $\mu$G (with weaker turbulences at smaller scales)
then the effects of the lens on $10^6$ GeV
cosmic rays can be observed, but from a displaced direction.
In any case, the identification of a CML would require
a detailed simulation including a full spectrum of magnetic 
turbulences.

We have studied the image of a point-like source, finding
interesting patterns that are the analogous
of the gravitational Einstein's ring and Einstein's cross.
Here the effect would be combined with a strong
{\it chromatic} dependence, as the deviation is proportional
to the inverse energy of the particle. The images would be
absent (or placed in a different location)
for particles of opposite charge, since they would
find a divergent lens.
We have also studied the effect of a CML on the
flux from a localized source.
If the source and the lens are far from the observer 
({\it i.e.}, if it covers a small solid angle) 
it seems possible to generate small-scale anisotropies. 
It would be interesting to study if under certain conditions
this type of fluctuations can survive into the 
diffuse regime (TeV cosmic rays) observed by 
Milagro\footnote{Milagro has also observed a large-scale 
anisotropy \cite{abdo2009} that could have 
a different origin \cite{battaner2010}.} 
\cite{abdo2008}.

\section*{Acknowledgments}
This work has been funded by MICINN of 
Spain (AYA2007-67625-CO2-02, FPA2006-05294, FPA2010-16802, 
and Consolider-Ingenio 
{\bf Multidark} CSD2009-00064) 
and by Junta de Andaluc\'\i a (FQM-101/108/437/792).


\begin{thebibliography}{99}

\bibitem{harari2001}  D.~Harari, S.~Mollerach and E.~Roulet,
AIP Conf.\ Proc.\ {\bf  566} (2001) 289. 

\bibitem{harari2005} Harari, D., 
  D.~Harari, S.~Mollerach and E.~Roulet,
  AIP Conf.\ Proc.\  {\bf 784} (2005) 763.

\bibitem{harari2010}  D.~Harari, S.~Mollerach and E.~Roulet,  
{\it Effects of the galactic magnetic field upon large scale
anisotropies of extragalactic cosmic rays} (2010) arXiv:1009.5891.

\bibitem{dolag2009} 
  K.~Dolag, M.~Kachelriess and D.~V.~Semikoz,
  JCAP {\bf 0901} (2009) 033
  [arXiv:0809.5055 [astro-ph]].

\bibitem{shaviv1999} 
  N.~J.~Shaviv, J.~S.~Heyl and Y.~Lithwick,
  Mon.\ Not.\ Roy.\ Astron.\ Soc.\  {\bf 306} (1999) 333 
  [arXiv:astro-ph/9901376].

\bibitem{beck2005} R.~Beck,  
{\it Cosmic Magnetic Fields}, p. 41. 
Springer Verlag, Ed. by R. Beck and R. Wielebinski.

\bibitem{parker1971}
E.~Parker, Astrophys. J. {\bf 163} (1971)  279.

\bibitem{brandenburg2005} 
  A.~Brandenburg and K.~Subramanian,
  Phys.\ Rept.\  {\bf 417} (2005) 1
  [arXiv:astro-ph/0405052].
 
\bibitem{lemaitre1933} G.~Lemaitre and M.S.~Vallarta, 
Phys.\ Rev.\ {\bf 43} (1933) 87.

\bibitem{battaner2008} E.~Battaner et al.  
in {\it Lecture Notes and Essays in 
Astrophysics}, Vol. 3, (2008) 83--102.

\bibitem{han2009} 
J.~L.~Han, in {\it Cosmic magnetic fields: from
planets to stars and galaxies,} IAU Conf.\ Proc.\ 259 (2009) 455
[arXiv:0901.1165 [astro-ph]].

\bibitem{ruiz2009} 
B.~Ruiz-Granados, J.A.~Rubi\~no-Mart\'\i n, E.~Battaner
in {\it Cosmic magnetic fields: from
planets to stars and galaxies,} IAU Conf.\ Proc.\ {\bf 259} (2009) 573.

\bibitem{dettmar2006} 
R.J.~Dettmar and M.~Soida, Astron.\ Nachr.\ {\bf 327} (2006) 495.

\bibitem{bonafede2009} 
  A.~Bonafede {\it et al.},
  Astron.\ Astrophys.\ {\bf 503} (2009)  707  
  [arXiv:0905.3552 [astro-ph.CO]].

\bibitem{han1997} 
J.L.~Han, R.N.~Manchester,  E.M.~Berkhuijsen,  R.~Beck, 
 Astron.\ Astrophys.\ {\bf 352} (1997)  98.

\bibitem{gonzalez1997} 
R.~González-Delgado, E.~Pérez, ApJ SS {\bf 108} (1997)  1.

\bibitem{crutcher2005} 
R.M.~Crutcher in {\it Magnetic Fields in the Universe: 
from Laboratory and Stars to the Primordial Structures}. Ed. by 
E.M. de Gouveia dal Pino et al. AIP Conf.\ Proc.\ 784 (2005) 129.

\bibitem{wielebinski2005} R.~Wielebinski 
in {\it The Magnetized Plasma in Galactic 
Evolution}. Proc. Conference. Ed. By K Chyzy et al. 
Jagiellonian Univ. Krakow. (2005) p. 125.

\bibitem{mitra2003} 
D.~Mitra, R.~Wielebinski, M.~Kramer and A.~Jessner, 
Astron.\ Astrophys.\ {\bf 398} (2003) 993.

\bibitem{stepinski1995} 
 T.F.~Stepinski, Rev. Mex. Astron. Astrophys., {\bf 1} (1995) 267.

\bibitem{goncalves2008} 
  J.~Goncalves, D.~Galli and J.~M.~Girart,
  Astron.\ Astrophys.\  {\bf 490} (2008) 39  [arXiv:0809.5278 [astro-ph]].

\bibitem{abdo2008} 
  A.~A.~Abdo {\it et al.},
  Phys.\ Rev.\ Lett.\  {\bf 101} (2008) 221101
  [arXiv:0801.3827 [astro-ph]].

\bibitem{abdo2009} 
  A.~A.~Abdo {\it et al.},
  Astrophys.\ J.\  {\bf 698} (2009) 2121
  [arXiv:0806.2293 [astro-ph]].
 
\bibitem{battaner2010} 
  E.~Battaner, J.~Castellano and M.~Masip,
  Astrophys.\ J.\  {\bf 703} (2009) L90 
  [arXiv:0907.2889 [astro-ph.HE]].


\end{thebibliography}
\end{document}